\title{  Time-domain inspiral templates for spinning compact binaries in quasi-circular orbits  }

\author{ Anuradha Gupta\footnote{Email: anuradha@iucaa.ernet.in: current address: 
Inter-University Centre for Astronomy and Astrophysics, Post Bag 4, Ganeshkhind, Pune 411 007, India} \, and Achamveedu Gopakumar\footnote{Email: gopu@tifr.res.in}\\ 
Department of Astronomy and Astrophysics
\\ Tata Institute of Fundamental Research
\\ Mumbai 400005,
India
}

\date{\today}

\documentclass[11pt,english]{article}
\usepackage{amsmath, amsthm, amssymb}
\usepackage[]{graphicx}

\newcommand{\vek}[1]{\boldsymbol{#1}}
\begin{document}
\maketitle

\begin{abstract}
We present a prescription to compute the time-domain gravitational wave (GW) 
polarization states associated with 
 spinning compact binaries inspiraling along quasi-circular orbits.
We invoke 
the orbital angular momentum $\vek L$
rather than its Newtonian counterpart $\vek L_{\rm N}$ to describe 
the orbits and the two spin vectors are freely specified 
in the source frame associated with the initial direction of the total
angular momentum. We discuss the various implications 
of our approach.

\end{abstract}

\section{Introduction}

Gravitational waves (GWs) from coalescing compact binaries containing at least one
spinning component are expected to be routinely detected by
the second-generation laser interferometric detectors like
advanced LIGO (aLIGO), Virgo and KAGRA \cite{RA}. 
The detection of GWs from such binaries and  subsequent 
source characterization crucially depend on accurately modeling
temporally evolving GW polarization states, $h_{+}(t)$ and $h_{\times}(t)$, from 
such binaries during their inspiral phase \cite{SS_lr}.
At present,  $h_{+}(t)$ and $h_{\times}(t)$  associated with non-spinning compact binaries 
inspiraling along quasi-circular orbits have GW phase evolution accurate to 3.5PN order 
and  amplitude corrections
that are 3PN accurate \cite{LB_lr}. Recall that the 3.5PN and 3PN orders correspond to 
corrections that are accurate to relative orders 
$(v/c)^7$ and $(v/c)^6$ beyond the `Newtonian' estimates, where $v$ and $c$ are the orbital 
and light speeds, respectively.
In the case of inspiraling compact binaries containing Kerr black Holes (BHs), 
it is desirable to employ temporally evolving  $h_{+}(t)$ and $h_{\times}(t)$
that incorporate the spin effects very accurately and the dominant spin effect arises  due to
the general relativistic spin-orbit coupling
that appear at the relative 1.5PN order for maximally spinning Kerr BHs \cite{BO,LK_95}.
Following Ref.~\cite{LK_95}, we define the spin of a compact object as
$ \vek S = G\, m_{\text {co}}^2\, \chi\, {\vek s}/c$, where
$m_{\text {co}}, \chi$ and $\vek s$ are its mass, Kerr parameter and 
a unit vector along $ \vek S$, respectively and for a maximally spinning Kerr BH $\chi=1$.
We note that it is customary to employ 
the Newtonian orbital angular momentum $\vek L_{\rm N} = \mu\, \vek r \times \vek v $,
where $\mu, \vek r$ and $\vek v$ are the reduced mass, orbital separation and 
velocity, respectively, to specify these quasi-circular orbits \cite{LK_95,ABFO}.

   In what follows we provide a prescription to generate the time-domain 
amplitude corrected  $h_{+}(t)$ and $h_{\times}(t)$ for spinning compact binaries
inspiraling along quasi-circular orbits, described by
their orbital angular 
momenta. Note that the amplitude corrected  $h_{+}(t)$ and $h_{\times}(t)$ refer to
GW polarization states that are PN-accurate both in its amplitude and phase.
We describe our approach to compute the fully 1.5PN accurate amplitude corrected 
expression for $h_{\times}(t)$ while invoking $\vek L$, the PN-accurate orbital angular momentum, to characterize the binary orbits.
We symbolically obtain the additional 1.5PN order amplitude corrections to $h_{\times}(t)$
in comparison with equations~(A3) in Ref.~\cite{ABFO} that employ $\vek L_{\rm N}$ to describe the binary orbits.
We also discuss certain implications of our approach while considering spin effects 
due to the leading order general relativistic spin-orbit coupling.
Our attempt to perform GW phasing
with the help of $\vek L$ is motivated by a number of observations
(we term accurate modeling of 
temporally evolving GW polarization states as `GW phasing').
 First 
being Ref.~\cite{GS11} that provided 
a prescription to implement 
GW phasing for spinning compact binaries in inspiralling
eccentric orbits in an accurate and efficient way.
We are further influenced by the fact that it is 
customary to use precessional equation appropriate for $\vek L$ 
to evolve $\vek L_{\rm N}$  while incorporating
the effects due to the dominant order spin-orbit coupling \cite{LK_95,ABFO,BCV}.
Finally, we note that a seminal paper that explored the inspiral
dynamics of spinning compact binaries and the influences of precessional 
dynamics on $h_{+, \times}(t)$ employed $\vek L$ to describe their binary 
orbits \cite{ACST}. Additionally, we specify the two spins in an inertial frame 
associated with the initial direction of the total angular momentum $\vek j_0$
 while it is customary to invoke a non-inertial $\vek L_{\rm N}$-based orbital triad 
to specify the two spins 
in the literature.

\section{GW phasing for spinning binaries characterized by $\vek L$}

\begin{figure}
\begin{center}
\includegraphics[height=6cm]{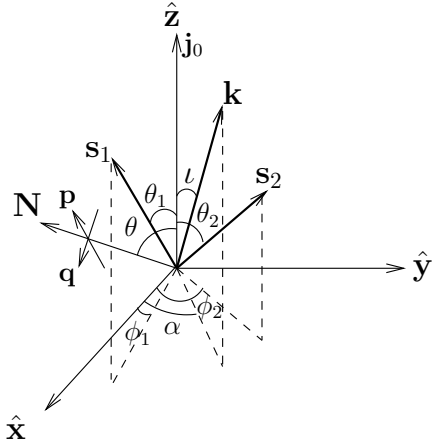}
\end{center}
\caption{ The inertial frame and the Cartesian coordinate system where 
the $\vek {\hat z}$ axis points along 
$\vek j_0$, the direction of total angular momentum at the initial epoch. We 
display the angles that characterize the orbital and spin angular momentum vectors, denoted 
by $\vek k, \vek s_1$ and $\vek s_2$, while the line of sight vector $\vek N$ is 
in the $x-z$ plane. The dashed lines depict the projections of various vectors on the $x-y$ plane.
}
\label{figure:frame}
\end{figure}

 We invoke, as noted earlier, the PN-accurate
orbital angular momentum $\vek L$ to describe
binary orbits and we specify at the initial epoch
both the orbital and spin angular momentum vectors 
in an inertial frame associated with the initial direction of 
total angular momentum (see figure~\ref{figure:frame}).
We begin by presenting an expression for 
the cross polarization state having Newtonian (quadrupolar)
order amplitude
and the relevant expression reads \cite{GG1}

\begin{align}
\label{eq:h+}
h_{\times}|_{\rm Q}(t) &= 2\, \frac{\, G\, \mu\,}{c^2\, R'}\, \frac{ v^2}{c^2}\Biggl\{ (1
-\cos \iota)\, S_{\theta}\,  \sin \iota\, \sin (\alpha- 2 \Phi)   \nonumber \\
&\quad - (1+\cos \iota )\,S_{\theta}\, \sin \iota \, \sin (\alpha + 2\Phi)  \nonumber \\
&\quad -\frac{1}{2}(1+ 2 \cos \iota+ \cos^2 \iota)\, C_{\theta} \sin (2\alpha +2\Phi)  \nonumber \\
&\quad -\frac{1}{2}(1- 2 \cos \iota+ \cos^2 \iota)\, C_{\theta} \sin (2\alpha -2\Phi)  
\Biggr\} \,,
\end{align}
where $R', S_{\theta}$ and $ C_{\theta}$ stand 
for the radial distance to the binary, $\sin \theta$ and $\cos \theta$, respectively.
The dynamical angular variable $\Phi $
measures the orbital phase from the direction of ascending node in a plane perpendicular to $\vek k$
while its derivative is required to define
 $v^2/c^2 = ( G\, m\, \dot \Phi/c^3)^{2/3}$. To obtain
temporally evolving  $h_{\times}|_{\rm Q}(t)$ associated with spinning 
compact binaries inspiraling along quasi-circular orbits, we pursue the 
following steps.
First, we specify how the Eulerian angles $\alpha$ and $\iota$ 
that specify the orientation of $\vek L$
vary under the 
conservative orbital dynamics.
This requires us to employ three 
differential equations that describe 
 the precessional dynamics of the orbital and spin angular momenta
and these differential equations contain 
an orbital like frequency $\omega $ \cite{LK_95}.
The conservative evolution for $ \Phi $ and $\dot \Phi$  are governed by the following
differential equation
\begin{equation}
\label{Eq_phidot}
 \dot \Phi = \omega - \cos \iota \, \dot \alpha \,,
\end{equation}
where the orbital like frequency $\omega $ is defined by the relation $\omega = v/r$
that connects $\omega$ to the orbital separation and velocity.
This PN-accurate equation arises from
the vectorial expression for the orbital velocity in a co-moving triad \cite{LK_95}.
It is convenient to introduce a dimensionless parameter
 $x \equiv ( G\, m\, \omega_{\rm orb} /c^3)^{2/3}$
such that the above equation reads 
$\dot \Phi = ( x^{3/2} / ( G\, m/c^3)) - \cos \iota\, \dot \alpha $.
Thereafter, we impose 
the effects of gravitational radiation reaction on these differential
equations by specifying how $x$, appearing in these  differential equations,
vary during the binary inspiral.

  It turns out that we require to solve simultaneously the  precessional equations for 
$\vek k, \vek s_1$ and $\vek s_2$,
the unit vectors along $\vek L, \vek S_1$ and $\vek S_2$,
 to specify how $\alpha $ and $\iota$
vary under the conservative dynamics as the precessional dynamics of $\vek L, \vek S_1$ and $\vek S_2$ 
are intertwined.
In figure~\ref{figure:frame}, we display various angles that 
specify the orientations of these three unit vectors in the above inertial frame. 
The Cartesian 
components of 
these 
 vectors are 
\begin{subequations}
\begin{align}
\vek k &= (\sin \iota\,\cos \alpha\, ,\,\sin \iota\,\sin
\alpha\, ,\,\cos \iota)\,,\\
\label{eq:s1_s2_1}
\vek s_1 &= \left ( \sin \theta_1\,\cos \phi_1, \sin \theta_1\,\sin \phi_1, \cos \theta_1 \right )\,,\\  
\label{eq:s1_s2_2}
\vek s_2 &= \left ( \sin \theta_2\,\cos \phi_2, \sin \theta_2\,\sin \phi_2, \cos \theta_2 \right )\,.
\end{align}
\end{subequations}
 The precessional equations for the above three unit vectors,
 extractable from Ref.~\cite{BO,LK_95}, read
\begin{subequations}
\label{Eq_dxy_ks1s2}
\begin{eqnarray}
\label{Eq_kdot}
{\dot {\vek k}} &=&  \frac{c^3}{Gm}\, x^3\,\Bigg\{ \delta_1\,q\, \chi_1\, 
\left( \vek s_1\times\vek k\right) \label{eq:kdot} 
+\frac{\delta_2}{q}\, \chi_2\, \left( \vek s_2\times\vek k\right) \Bigg\} \,, \\
{\dot {\vek s}_{1}} &=&  \frac{c^3}{Gm}\, x^{5/2}\,\delta_1 \left(\vek k\times \vek s_1\right) \,, \\
{\dot {\vek s}_{2}} &=&  \frac{c^3}{Gm}\, x^{5/2}\,\delta_2 \left(\vek k\times \vek s_2\right) \,,
\end{eqnarray}
\end{subequations} 
where
$q=m_1/m_2$ and the 
symmetric mass ratio $\eta = \mu/m$ is required to define 
the quantities $\delta_1$ and $ \delta_2$ ( $\delta_{1,2}=\eta/2+3(1\mp \sqrt{1-4\eta})/4$).
We are now in a position to incorporate the effect of gravitational radiation damping
and this is achieved by specifying the secular variations in $x$ in the above 
differential equations.
The PN-accurate differential equation for $x$ arises from the usual energy balance
arguments and may be obtained from equation~(3.21) in Ref.~\cite{ABFO}.

 We obtain numerically the time-domain  GW polarization states for our binaries
by simultaneously solving the differential equations for the Cartesian components of
 $\vek k, \vek s_1$ and 
$\vek s_2$ along with the PN-accurate equations for $ \dot \Phi $ and $d x/dt$.
This implies that we numerically solve a set of $11$ differential equations 
and the resulting variations in  $\alpha, \iota, \Phi$ and $ \dot \Phi $ are implemented in 
equation~(\ref{eq:h+}) to obtain the temporally varying $h_{\times}|_{\rm Q}(t)$.
A close inspection reveals that 
we require to specify
four angles
that provide the orientations of the two spins from $\vek j_0$ at the initial epoch
to obtain $h_{\times}|_{\rm Q}(t)$.
It should be noted that the two angles that specify the  initial  orientation of $\vek k$
from $\vek j_0$ are not independent variables. This is 
because one may point 
the total angular momentum  along the $z$-axis at the initial epoch 
without the loss of any generality.
This allowed  us to equate the $x$ and $y$ components of $\vek J= \vek L + \vek S_1 + \vek S_2$ to zero at that epoch
and therefore to obtain the initial estimates for $\iota$ and $\alpha$.
In our numerical integrations, 
the bounding values for $x$ are
given by $x_0=2.9\times 10^{-4}\, (m'\, \omega_0)^{2/3} $ 
and $x_{\rm f}=1/6 $, where $m'$ is the total mass of the binary in solar units
and let $\omega_0 = 10\, \pi$ Hz as customary for aLIGO 
along with $\Phi(x_0) =0$.
Finally, we note that the values of $\alpha$ and $\iota$
at every step of our numerical runs are obtained from 
the Cartesian components of $\vek k$ by the relations:
$ \alpha = \cos^{-1}(k_{\rm x}/\sqrt{k_{\rm x}^2 + k_{\rm y}^2}) $ and $\iota = \cos^{-1} (k_{\rm z})$.
Therefore, we require eight independent parameters to specify the inspiral dynamics
of compact binaries with spinning components and these
eight  parameters are the four basic (constant) parameters 
namely $(m_1, m_2, \chi_1, \chi_2)$ along with the above four angular dynamical variables that specify the two
spin vectors in the inertial frame, namely $( \theta_1, \phi_1) $ and $( \theta_2, \phi_2) $ as displayed in figure~\ref{figure:frame}.

  In what follows we compare our approach with what is detailed in Ref.~\cite{ABFO}
that provided a way to obtain the time-domain ready-to-use GW polarization states
for inspiraling compact binaries while incorporating all 
1.5PN order spin effects both in amplitude and GW phase evolutions.
Their approach differs from ours in two 
aspects.  
First, Ref.~\cite{ABFO} employed 
$\vek l = \vek L_{\rm N}/| \vek L_{\rm N} |$ 
to characterize the orbital plane and following the 
earlier papers employed a precessional equation for $\vek l$ 
that is identical to our equation~(\ref{eq:kdot})
 for $\dot {\vek k}$ 
while replacing $\vek k$ by $\vek l$.
It is not very difficult to show that this is equivalent
of using an 
orbital averaged expression for $\dot{\vek l}$ \cite{LK_95}.
However, invoking an orbital averaged precessional equation for $\vek l$ 
leads to an undesirable feature that the coefficient of
$\vek l$ in the expression for $\dot {\vek n}$ in the 
$( {\vek n},{\vek {\lambda}} = \vek l \times \vek n,{\vek l} )$
frame will not, in general, vanish.
It is fairly straightforward to compute the time derivative of $\vek n$ 
and express it in the $( {\vek n},{\vek {\lambda}},{\vek l} )$
frame as
\begin{align}
\label{eq:dndt}
\frac{d {\vek n} }{dt} =  \biggl ( \frac{ d \Phi'}{dt} + \cos \iota' \,
 \frac{ d \alpha'}{dt} \biggr ) \, \vek \lambda
+ \biggl (  \frac{ d \iota' }{dt}\, \sin \Phi' 
- \sin \iota' \, \cos \Phi'\,  \frac{ d \alpha'}{dt} \biggr )\, \vek l
\,,
\end{align}
where $\iota'$, $\alpha'$ and $\Phi'$ are the usual 3 Eulerian angles
required to define $\vek n$, $\vek \lambda$ and $\vek l$ in the $\vek j_0$-based
inertial frame. 
The fact that $\vek l \equiv \vek n \times \dot{\vek n} /|\vek n \times \dot{\vek n}|$ 
clearly demands 
that the coefficient of $\vek l$ in the above equation should be zero 
as also noted in Ref.~\cite{ABFO}.
However, if one employs equation~(\ref{eq:kdot}) 
to describe the precessional dynamics of $\vek l$ 
then it is possible to show with some straightforward algebra that 
\begin{align}
\label{Eq_relation}
 \sin \Phi' \, \frac{d\iota'}{dt} - \cos \Phi' \, \sin \iota' \, \frac{d\alpha'}{dt} 
= -\frac{c^3}{G\, m}\, x^3\, \Bigg\{ \delta_1 \, q\, \chi_1\, (\vek s_1 \cdot \vek \lambda ) 
 + \frac{\delta_2}{q}\, \chi_2\, (\vek s_2 \cdot \vek \lambda)\Bigg\} \,,
\end{align}
as noted in Ref.~\cite{GS11}.
It is not very difficult to conclude that 
the right hand side of above expression, in general, is not zero.

It turns out that $\dot {\vek n}$ having components along $\vek l$ 
leads to certain anomalous terms that contribute to the $\Phi'$ evolution at the 3PN order
and this is within the consideration of higher order spin effects  
available in the literature. 
Note that the higher order spin effects are known to 3.5PN order while 
dealing with the spin-orbit interactions \cite{MBFB_2012}. 
We demonstrate our observation by noting that the definitions $ \vek v = r\, \dot {\vek n}$ and 
$ v = r\, \omega $ imply that $ \omega^2 = \dot {\vek n} \cdot \dot {\vek n}$.
With the help of our Eq.~(\ref{eq:dndt}) the expression for $ \omega$ reads

\begin{align}
 \omega = \biggl ( \frac{ d \Phi'}{dt} + \cos \iota' \,
 \frac{ d \alpha'}{dt} \biggr ) 
+  \frac{1}{2\, \dot \Phi'} \biggl (  \frac{ d \iota' }{dt}\, \sin \Phi' 
- \sin \iota' \, \cos \Phi'\,  \frac{ d \alpha'}{dt} \biggr )^2\,,
\end{align}
where $ \dot \Phi'$ stands for $ d \Phi'/{dt}$ and the higher order terms 
that are cubic in the time derivatives of $\iota'$ and $\alpha'$ are neglected.
Invoking the fact that $\dot \Phi'$ at the Newtonian order is given by $ x^{3/2}/ (G\,m/c^3)$
and noting that the expression for $d \iota'/dt $ and $d \alpha'/dt $ arise from the 1.5PN order differential equation for $\dot{\vek k}$,
we get
\begin{align}
\dot { \Phi'}&= \frac{c^3}{G\,m}\, x^{3/2} \biggl ( 1 + x^{3/2} \, A' + x^{3} \, B'\biggr ) \,,
\end{align}
such that $A'$ and $B'$ are given by
\begin{subequations}
 \begin{eqnarray}
 A' &= -\frac{\cos \iota'}{\sin \iota'}\Big \{ [\delta_1 \, q\, \chi_1\, (\vek s_1 \cdot \vek \lambda )
+ \frac{\delta_2}{q}\, \chi_2\, (\vek s_2 \cdot \vek \lambda)] \, \cos \Phi'  \nonumber \\
&\quad+ [\delta_1 \, q\, \chi_1\, (\vek s_1 \cdot \vek n )    
+ \frac{\delta_2}{q}\, \chi_2\, (\vek s_2 \cdot \vek n)]\, \sin \Phi'\Big \} \,, \\ 
B' &= -\frac{1}{2}\, \Big \{ \delta_1 \, q\, \chi_1\, (\vek s_1 \cdot \vek \lambda )
+ \frac{\delta_2}{q}\, \chi_2\, (\vek s_2 \cdot \vek \lambda) \Big \}^2 \,.
\end{eqnarray}
\end{subequations}
It should be evident that the  $B'$ terms appear at the 3PN order 
while $A'$ terms enter $\dot{\Phi}$ expression at the 1.5PN order.
It is not very difficult to infer that these  $B'$ terms arise due to the non-vanishing
$\vek l$ component in the expression for $\dot {\vek n}$
and therefore are unphysical in nature.
Therefore, these anomalous terms contribute to the $\Phi'$ evolution at the third 
post-Newtonian order and this is within the consideration
of higher order spin effects currently available in the literature.
It should be noted that these unphysical terms play no role in investigations in Ref.~\cite{ABFO}
that probed the leading order spin effects appearing at the 
1.5PN order in the phase evolution.

  Another consequence of invoking $\vek k$ to specify the binary orbit is the appearance of 
certain new 1.5PN order contributions to the amplitudes of $h_{+} $ and $h_{\times}$
in addition to what is provided by equations~(A1), (A2) and (A3) in Ref.~\cite{ABFO}.
We recall that equations~(A1), (A2) and (A3) in Ref.~\cite{ABFO} provide the fully 1.5PN 
accurate expressions for $h_{+} $ and $h_{\times}$ while invoking 
$\vek L_{\rm N}$ to describe the binary orbits.
The additional amplitude corrections to the GW polarization states arise 
mainly due to the fact
 that the component of $\vek v$ along $\vek k$ is of 1.5PN order.
To demonstrate this point,  
we  express $\vek r$ and $\vek v = d \vek r/dt $  in the inertial frame 
$\left ( \hat {\vek x}, \hat {\vek y}, \hat {\vek z} \right )$
associated with $\vek j_0$ 
with the help of  
the three usual Eulerian angles $\Phi, \alpha$ and  $\iota$ as displayed in figure~\ref{figure:frame}.
The relevant expression for $\vek r $ may be written as
$\vek r = r \vek n $ where $\vek n = 
(-\sin \alpha\,\cos \Phi-\cos \iota\,\cos \alpha\,\sin 
\Phi)\hat {\vek x} + (  
 \cos \alpha\,\cos \Phi   
-\cos \iota\,\sin \alpha\,\sin \Phi ) \hat {\vek y}
+  \sin \iota\,\sin \Phi \hat {\vek z} $. 
To show that $\vek v$ can have non-vanishing 1.5PN order terms along $\vek k$,
we compute $ d \vek r/dt $ 
in the co-moving frame defined by the 
triad $( {\vek n},{\vek {\xi}} = \vek k \times \vek n,{\vek k} )$
and this is easily achieved with the help of three rotations involving the 
three Eulerian angles appearing in the expression for $\vek r $ \cite{DS_88}.
The resulting expression for $\vek v$ reads
\begin{align}
 \vek v &=  
r\, \biggl (  \frac{ d \Phi}{dt} +   \frac{ d \alpha}{dt}\,
\cos \iota
 \biggr ) \, \vek \xi
+ r\, \biggl (  \frac{ d \iota }{dt}\, \sin  \Phi 
- \sin \iota \, \cos \Phi\,  \frac{ d \alpha}{dt} \biggr )\, \vek k
\,. 
\end{align}
It is not very difficult to verify that $\vek v \cdot \vek k \ne 0$ 
while invoking equation~(\ref{Eq_kdot}) for $ \dot {\vek k}$
to evaluate the $\vek k$ component of $\vek v$.
Moreover, the coefficient of  $\vek k$ in the above expression
 for $\vek v$ is at the 1.5PN order.
 The $\vek k$ component of $\vek v$ enters the 
expressions for $h_{+} $ and $h_{\times}$ through the 
dot products $( \vek p \cdot \vek v )\, ( \vek q \cdot \vek v ) $ and 
$( \vek p \cdot \vek v )^2 - ( \vek q \cdot \vek v ) ^2 $
that are required to compute the PN-accurate expressions for $h_{+} $ and $h_{\times}$
(the vectors $\vek p$ and $\vek q$ are two unit vectors that lay in a plane 
perpendicular to $\vek N$).
The resulting 1.5PN order amplitude corrections to $h_{\times}$ symbolically
read
\begin{align}
\label{hc15}
h_{\times}\Big|_{1.5\rm PN} &= \frac{G\, \mu}{c^4\, R'}\, \frac{G\, m}{c^3\, \sqrt{x}}\,
 \Biggl\{  \frac{d {\iota}}{dt}\, \kappa_1(\iota, \alpha, \Phi, \theta)+ 
\frac{d \alpha}{dt}\, \kappa_2(\iota, \alpha, \Phi, \theta) \Biggr \}  \,,
\end{align}
where the explicit expressions for $\kappa_1$ and $\kappa_2$ are available in Ref.~\cite{GG1}. 
Therefore, the fully 1.5PN order amplitude corrected expression for $h_{\times}$
associated with the spinning compact binaries 
in quasi-circular orbits, described by $\vek L$, is 
provided by equations~(A3) along with the above equation~(\ref{hc15}).
This statement also requires that the angular variables $\iota$ and $\alpha$
that appear in equations~(A3) of  Ref.~\cite{ABFO} represent $\vek k$ rather than $\vek l$.
Let us emphasize again that  equations~(A2) and (A3) in Ref.~\cite{ABFO} indeed
provide the fully 1.5PN accurate amplitude corrected $h_{+} $ and $h_{\times}$
for spinning compact binaries in circular orbits described by $\vek L_{\rm N}$.

 Another aspect where we differ from
 Refs.~\cite{ABFO,BCV} is the way we specify the two
spin and $\vek k$ vectors to perform GW phasing.
In the literature, it is common to freely specify spin vectors in an orthonormal
triad defined by using $ \vek l$ \cite{LK_95,BCV,ABFO}.
In contrast, we freely specify the two spin vectors at the initial epoch in the inertial source
 frame associated with $\vek j_0$.
This choice allowed us to specify the $x$ and $y$
components of $\vek k$ 
uniquely in terms of $\vek s_1, \vek s_2$ and other intrinsic binary parameters at the initial epoch
by demanding that the
$x$ and $y$ components of $\vek J= \vek L + \vek S_1 + \vek S_2$ should be zero at that epoch.
Therefore, we extract very easily 
the initial values of $\alpha$ and $\iota$ from the following 
expressions for the initial $x$ and $y$
components of $\vek k$ 
\begin{subequations}
\label{Eq_ialpha_ini}
\begin{eqnarray}
k_{\rm x,i}=\sin \iota\, \cos \alpha&=&-\frac{G\, m^2}{c\,L_{\rm i}}\,
\{X_1^2\, \chi_1\,\sin \theta'_1\,\cos \phi'_1  
+X_2^2\, \chi_2\,\sin \theta'_2\,\cos \phi'_2\} \,, \nonumber \\
k_{\rm y,i}=\sin \iota\, \sin \alpha&=&-\frac{G\, m^2}{c\,L_{\rm i}}\,
\{X_1^2\, \chi_1\,\sin \theta'_1\,\sin \phi'_1   
+X_2^2\, \chi_2\,\sin \theta'_2\,\sin \phi'_2\} \,, \nonumber
\end{eqnarray}
\end{subequations}
where $\theta_1', \phi_1', \theta_2', \phi_2'$ are the values of $\theta_1, \phi_1, \theta_2, \phi_2$
at the initial epoch and
$L_{\rm i}$ denotes the PN-accurate expression for 
$|\vek L|$ at initial orbital frequency.

 For isolated unequal mass spinning compact binaries with $q>3$ it should be advantageous to 
specify their spins in the inertial frame associated with $\vek j_0$.
This is because $\theta_1(x_0)$ that provides 
the dominant spin orientation from $\vek j_0$ at $x_0$ is expected to lie 
in a smaller range for 
such binaries spiraling into $x_0$ due to the emission of GWs.
A recent study reveals that the dominant BH spin orientation from $\vek j_0$ at $x_0$
is more likely to be $\leq 90^{\circ} $ for angular momentum dominated unequal mass binaries
$( |\vek L|(x_0) > S_1)$
while $\theta_1(x_0) \leq 45^{\circ} $ for unequal mass binaries having $S_1 > |\vek L|$ at $x_0$ \cite{GG2}.
These inferences require that 
these binaries inherit spin-orbit misalignments $\leq 160^{\circ} $ from
various astrophysical processes responsible for the formation of such unequal mass spinning binaries.
The physical explanation for these conclusions is the
observed  alignment of $\vek S_1$ towards $\vek j_0$ 
due to the action of gravitational radiation reaction, detailed in Ref.~\cite{ACST}.
However, it will be difficult to provide similar bounds for the dominant spin orientation 
if the spins are freely specified in a non-inertial orbital triad  associated with 
$\vek L$ at $x_0$. 
In this case  $\vek k \cdot \vek s_1$ provides the dominant spin orientation 
at $x_0$ and it is not difficult to show that $\vek k \cdot \vek s_1$ remains 
fairly constant as these unequal mass binaries with $q>3$ spiral into $x_0$
from initial orbital separations $\leq 1000\, G\, m/c^2$ due to the emission of GWs.



\end{document}